\documentclass[traditabstract]{aa}  
\usepackage{natbib} 
\bibpunct{(}{)}{;}{a}{}{,} 
\usepackage{graphicx}
\usepackage{txfonts}
\usepackage{amssymb}
\def\inte{{\em INTEGRAL}}

\def\chan{{\em Chandra}}

\def\swift{{\em Swift}}

\def \inte {{$INTEGRAL$}}

\def \chandra {{$Chandra$}}

\def \igr {\mbox{IGR\,J18179-1621}}

\def \ferg {erg~cm$^{-2}$~s$^{-1}$}
\def \hcm {\hbox {\ifmmode $ atom cm$^{-2}\else atom cm$^{-2}$\fi}}
\def \arcmin {\hbox{$^\prime$}}

\def \apjl {ApJL}
\def \apjs {ApJS}\def \aap {A\&A}

\begin{document}
   \title{ \igr:\ An obscured X-ray pulsar discovered by \inte\ }

\author{E. Bozzo 
   \inst{1}
	\and 
   C. Ferrigno 
 	\inst{1} 
 	\and   
   M. T\"urler  
   \inst{1}  
   \and 
   A. Manousakis
   \inst{1}
   	\and	
	M. Falanga
		\inst{2}		
}

\authorrunning{E. Bozzo et al.}
  \titlerunning{ \igr:\ an obscured X-ray pulsar discovered by \inte\ }
  \offprints{enrico.bozzo@unige.ch}

\institute{ISDC, Data Center for Astrophysics of the University of Geneva, 
	 chemin d'\'Ecogia, 16 1290 Versoix Switzerland 
         \and
        International Space Science Institute (ISSI), Hallerstrasse 6, CH-3012 Bern, Switzerland    
         }

 \abstract{We report on all the \inte\ and \swift\ data collected  during the first outburst observed from  
 \igr.\  The broad-band spectral analysis 
 showed that the X-ray emission from the source is heavily absorbed 
 ($N_\mathrm{H}\simeq$10$^{23}$~cm$^{-2}$), and well-described by a flat power-law with a high energy rollover 
 (cutoff energy 9-12~keV, e-folding energy 4-7~keV). We found some evidence of 
 a cyclotron absorption feature at $22\pm1$~keV. Together with the pulsations at 11.8~s discovered in the XRT data, this evidence would suggest 
 that \igr\ is an obscured, magnetized, accreting neutron star that is possibly part of a supergiant high-mass X-ray binary or a Be X-ray binary system.}   
   

   \date{Received: 2012 April 5; accepted: 2012 July 19}

   \maketitle

\section{Introduction}
\label{sec:intro} 

\igr\ was discovered by the hard X-ray imager ISGRI on-board 
\inte\ on 2012 February 29 \citep{turler12}.  
The estimated source flux in the 20-40 keV energy band 
was 16$\pm$1~mCrab, $(1.23\pm0.08)\times10^{-10}$~\ferg. 
Follow-up observations with \swift\ and \chan\ permitted 
to obtain the refined position of the source at $\alpha_{\rm J2000}$=18$^{\rm h}$17$^{\rm m}$52$\fs$19 and 
$\delta_{\rm J2000}$=-16${\degr}$21$\arcmin$31$\farcs$7 (error-circle radius
of 0$\farcs$.6 at 90\% confidence level), and led to the identification of the candidate infrared 
(IR) counterpart 2MASS\,J18175218$-$1621316
\citep{li12,paizis12}.  
The narrow-field instrument on-board \swift,\ XRT, also measured a high absorption column-density in the 
direction of the source ($\sim$10$^{23}$~cm$^{-2}$) and detected  
pulsations in its X-ray emission at a period of 11.82~s \citep{turler12,halpern12}, 
which was confirmed using FERMI/GBM \citep{finger2012}.
This evidence suggested that \igr\ is a highly obscured X-ray pulsar, possibly part of 
a high-mass X-ray binary system. 

In this paper, we report on the analysis of all the \inte\,/ISGRI and \swift\,/XRT target-of-opportunity observations we requested 
to be performed in the direction of \igr,\ and discuss different interpretations of the nature of the source.

\section{Data analysis and results}
\label{sec:data}

\subsection{ \inte\ }
\label{sec:integral}

\igr\ was detected simultaneously by IBIS/ISGRI \citep[15-500~keV,][]{ubertini03,lebrun03} 
and JEM-X \citep[3-25~keV,][]{lund03} for the first time during satellite revolution 1145, and observed throughout 
revolutions 1146 and 1147 (from MJD~55\,986.1 to MJD~55\,992.5). 
Further observations were inhibited by a bright solar flare 
that forced the instrument on-board \inte\ to enter in ``safe-mode'' with no scientific data 
available\footnote{See http://www.isdc.unige.ch/integral/operations/reports}.    
A complete log of these observations is provided in Table~\ref{tab:integral}.  
\begin{table}
\centering
\scriptsize
\caption{ \inte\ observation log and effective exposure times of the spectra of \igr.\ } 
\begin{tabular}{@{}llllll@{}}
\hline
\hline
\noalign{\smallskip}
Rev.  & START TIME & STOP TIME & ISGRI  & JEM-X1 & JEM-X2 \\
      &    (MJD)   &   (MJD)   & (ks) & (ks) & (ks) \\
\noalign{\smallskip}
\hline
\noalign{\smallskip}
1145       & 55986.1 & 55987.8 & 55.6 & 23.7 & 23.1\\
\noalign{\smallskip}
1146 & 55988.9 & 55990.8 & 62.8 & 35.3 & 34.6\\
\noalign{\smallskip}
1147 & 55991.8 & 55992.5 & 37.4 & 21.8 & 20.9 \\
\noalign{\smallskip}
\hline
\noalign{\smallskip}
\multicolumn{6}{l}{Note: A selection for source off-set angles has been applied, see Sect.~\ref{sec:integral}.} \\
\end{tabular}
\label{tab:integral}
\end{table} 
All \inte\ data were analyzed using version 9.0 of the OSA 
software distributed by the ISDC \citep{courvoisier03}. 
\inte\ observations are commonly divided into ``science windows'' (SCWs), 
i.e. pointings with typical durations of $\sim$2-3~ks.  
Because of uncertainties in the instrument responses for high off-axis angles,
we selected for the spectral analysis only those SCWs for which the off-set angle of the source was 
$<$3.5~deg and $<$12~deg, respectively, for JEM-X and ISGRI. The total effective exposure-time 
resulting after this selection is reported in Table~\ref{tab:integral}. 
In all revolutions, the source was relatively faint for the JEM-X sensitivity, thus we performed a spectral extraction by using the option {\it flag=1} 
in the local source catalog\footnote{This means that the source spectrum is extracted at the given catalog position even if the source is not detected 
within a single science window, see http://www.isdc.unige.ch/integral/download/osa/doc/current/osa\_um \_jemx/node40.html}. 
We checked {\it a posteriori} that this did not affect the 
discussion on the main spectral results reported in Sect.~\ref{sec:swift}. 
Owing to the relatively low signal-to-noise ratio (S/N) of the \inte\ data, no meaningful timing analysis could be carried out.

\subsection{ \swift\ }
\label{sec:swift}

\swift\,/XRT observations were performed from a few hours up to $\sim$22~days following 
the discovery of the source (29 Feb 2012-- 22 Mar 2012, see Table~\ref{tab:swift}). 
\swift\,/XRT data were collected only in photon-counting (PC) mode and 
analyzed by using standard procedures \citep{burrows05}.  
The XRT data were processed with the {\sc xrtpipeline} 
(v.0.12.6) and the latest calibration files available \citep[see also][]{bozzo09}. 
When required, we corrected PC data for pile-up, and used the 
{\sc xrtlccorr} to account for this correction in the 
background-subtracted light curves. Source and background-event lists (time resolution 2.5~s) 
were barycentered by using the {\sc barycorr} tool.  
\begin{table}
\centering
\scriptsize
\caption{ \swift\ observations log of \igr.\ } 
\begin{tabular}{@{}llllllll@{}}
\hline
\hline
\noalign{\smallskip}
OBS$^a$ & START$^b$ & STOP$^b$ & Exp & $N_{\rm H}$$^c$ & $\Gamma$ & $F_{\rm obs}$$^d$ & $\chi^2_{\rm red}$/d.o.f. \\
       &     (day)       &   (day)        &  (ks)   &  & & & (C-stat/d.o.f.) \\
\noalign{\smallskip}
\hline
\noalign{\smallskip}
01 & 0.85  & 0.92  & 2.0 & 10.9$^{+4.7}_{-4.1}$ & 0.1$\pm$0.5 & 2.9 & 0.6/31 \\
\noalign{\smallskip}
02 & 1.64  & 1.67  & 1.9 & 17.7$^{+6.7}_{-5.8}$ & 0.6$^{+0.7}_{-0.6}$ & 2.8 & 1.0/25 \\
\noalign{\smallskip}
04 & 2.92  & 3.00  & 2.3 & 13.3$^{+5.8}_{-5.1}$  & 0.3$^{+0.6}_{-0.6}$ & 2.6 & 1.0/23 \\
\noalign{\smallskip}
05 & 3.32  & 3.47  & 2.2 & 9.2$^{+5.1}_{-4.4}$  & 0$^{+0.5}_{-0.5}$ & 2.8 & 0.8/26 \\
\noalign{\smallskip}
06 & 4.52  & 4.74  & 2.2 & 14.7$^{+4.8}_{-4.2}$  & 0.2$^{+0.5}_{-0.5}$ & 2.0 & 1.2/37 \\
\noalign{\smallskip}
07 & 6.12  & 6.40  & 1.9 & 9.2$^{+5.8}_{-5.1}$  & 0$^{+0.6}_{-0.6}$ & 1.6 & 1.2/25 \\
\noalign{\smallskip}
08 & 7.33  & 7.82  & 2.2 & 12.6$^{+4.7}_{-4.1}$  & 0$^{+0.5}_{-0.5}$ & 1.3 & 0.9/35 \\
\noalign{\smallskip}
09 & 9.22 & 10.23  & 2.2 & 16.6$^{+4.5}_{-4.0}$  & 0.6$^{+0.5}_{-0.5}$ & 1.0 & 1.0/37 \\
\noalign{\smallskip}
10 & 11.21  & 11.45  & 1.3 & 10.9$^{+7.6}_{-6.9}$  & 0.3$^{+0.8}_{-0.7}$ & 0.87 & 1.1/19 \\
\noalign{\smallskip}
12 & 13.03  & 13.70  & 1.5 & 10.4$^{+8.0}_{-7.0}$  & 0.4$^{+0.9}_{-0.8}$ & 0.54 & 0.9/14 \\
\noalign{\smallskip}
13 & 17.63  & 17.64  & 1.3 & 15.4$^{+17.7}_{-9.4}$  & 1.7$^{+2.4}_{-1.7}$ & 0.19 & (19.7/22) \\
\noalign{\smallskip}
14 & 18.30  & 18.31  & 1.1 & 18.3$^{+20.8}_{-13.5}$  & 0.7$^{+2.2}_{-1.8}$ & 0.23 & (14.5/17) \\
\noalign{\smallskip}
15-17$^e$ & 19.31  & 21.79  & 3.1 &  25.4$^{+11.8}_{-10.0}$ & 3.0$^{+1.5}_{-1.8}$ & 0.08 & (28.1/29) \\
\noalign{\smallskip}
\hline
\noalign{\smallskip}
\multicolumn{8}{l}{Note: all data were fit with a simple absorbed power-law model.} \\
\multicolumn{8}{l}{$^a$: Only the last two digits of the observation ID.000322930** are shown.} \\
\multicolumn{8}{l}{$^b$: Start and stop times of the observations are from MJD~55\,986, 29 Feb 2012.} \\ 
\multicolumn{8}{l}{$^c$: The absorption column density is in units of 10$^{22}$~cm$^{-2}$.} \\
\multicolumn{8}{l}{$^d$: The absorbed X-ray flux is in units of 10$^{-10}$\,\ferg}\\
\multicolumn{8}{l}{$^e$: Owing to the relatively low statistics, the three observations 15, 16, and 17 were} \\ 
\multicolumn{8}{l}{merged to allow for a meaningful spectral extraction.} \\
\end{tabular}
\label{tab:swift}
\end{table} 
\begin{figure}
\centering
\includegraphics[width=5.8cm,angle=-90]{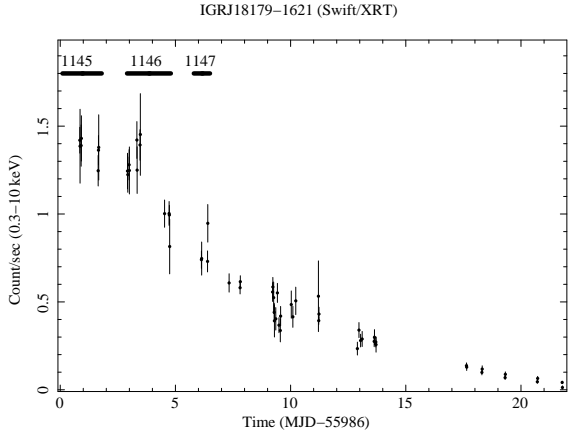}
\caption{ \swift\,/XRT lightcurve of \igr\ during the $\sim$22~days monitoring reported in this paper (time bin is 1~ks). 
The time intervals covered quasi-simultaneously by \inte\ is also reported with a solid line.}  
\label{fig:lcurve} 
\end{figure} 

\begin{figure}
\centering
\includegraphics[width=7cm,angle=-90]{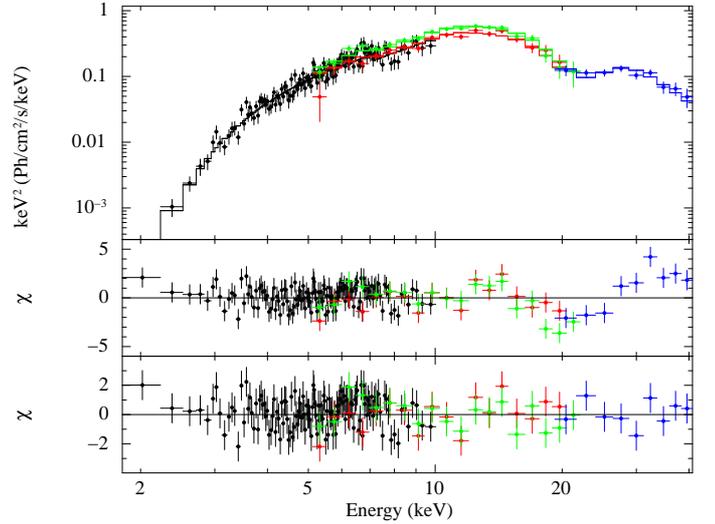}
\caption{Our XRT/JEM-X/ISGRI combined spectra during the \inte\ revolutions 1145 and 1146 (see text for details). 
The best-fit model is obtained by using a power law with a high-energy cut-off plus a cyclotron absorption-line 
at $\sim$20~keV. The mid panel shows the residuals from the fit when the cyclotron feature is not included.}  
\label{fig:spectra} 
\end{figure}

All XRT spectra were well-described by a power-law model with a relatively large absorption,
$N_{\rm H}$$\gtrsim$10$^{23}$~cm$^{-2}$ (we used the {\sc phabs} model in {\sc Xspec}12),   
and flat photon index ($\Gamma$$\sim$0; see Table~\ref{tab:swift}). In order to obtain a better estimate of the source X-ray spectral properties, we performed a quasi-simultaneous 
fit of the XRT, JEM-X and ISGRI data. XRT observations from 00032293001 to 00032293006 were summed-up together 
to perform the quasi-simultaneous fit with the JEM-X and ISGRI data in revolutions 1145 and 1146, where the source displayed only a moderate change in the X-ray flux 
(we verified that no significant spectral variation occurred in the two revolutions by performing separate fits to the data). 
As the source flux during revolution 1147 is significantly lower than during the previous one,
we performed a separate fit by combining the XRT spectrum obtained from the observation 00032293007 with the JEM-X and ISGRI spectra extracted during revolution 1147. 
The results of this analysis are shown in Fig.~\ref{fig:spectra} and reported in Table~\ref{tab:results}. 

Spectra from the first data-set (1145+1146) were fit first using a number of phenomenological models, including in all cases a normalization 
constant to take into account the systematic uncertainty in the inter-calibrations among the instruments and the possible variability of the source (the constant was fixed to 1 for XRT). 
A single absorbed power-law model failed to provide an acceptable fit to the spectra ($\chi^2_{\rm red}$$\simeq$3.5).  
The addition of a cut-off at the higher energies ({\sc highecut*pow} in {\sc Xspec}) significantly  
improved the fit ($\chi^2_{\rm red}$$\simeq$1.5), but left evident residuals above $\sim$10~keV (see Fig.~\ref{fig:spectra}). 
The widely used NPEX and Fermi-Dirac cut-off models \citep[see e.g.,][]{coburn2002} did not give a better fit ($\chi^2_\mathrm{red}\simeq1.7$) and left very similar residuals.  
We also tried the most widely used Comptonization models: {\sc BMC} comprises black-body (BB) 
radiation up-scattered by a spherically symmetric in-falling plasma
\citep{shrader98} and provided a slightly better fit ($\chi^2_{\rm red}$=1.4). However, 
the temperature of the thermal seed photons was significantly higher ($\sim$3~keV) than expected in the case of an accreting 
neutron star \citep[0.8-1.5~keV; see e.g.,][]{orlandini06}. Similarly, a {\sc comptt} model \citep{comptt}, which accounts for the 
thermal Comptonization of BB seed photons, was bounded to a temperature of the BB $\ga$2.5\,keV.
In all models, evident residuals were left above $\sim$10~keV. 
As suggested by the preliminary data analysis reported by \citet{turler12}, the combination of a cut-off power-law model plus a cyclotron 
absorption line at $\sim$20~keV gave a reasonably good fit to the data. However, the cyclotron feature parameters could only be poorly constrained; we thus 
fixed the line width ($\sigma_{\rm cycl}$) to 3\,keV, which is a reasonable value for an X-ray pulsar 
\citep[see, e.g.,][and Table~\ref{tab:results}]{ferrigno09}.   
In revolution 1147, the S/N of the data was relatively low compared to data in revolutions 1145 and 1146, and 
the possible cyclotron feature was not formally required in the fit (an acceptable fit could be obtained by using an absorbed  
power-law model with a high energy cut-off, see Table~\ref{tab:results}). 
A further discussion of the interpretation of these results is given in Sect.~\ref{sec:discussion}. 
\begin{table*}
\centering
\scriptsize
\caption{Results of the combined XRT/JEM-X/ISGRI spectra of \igr.\  using an absorbed power-law with high-energy exponential roll-over and an absorption feature.} 
\begin{tabular}{@{}llllllllllllllll@{}}
\hline
\hline
\noalign{\smallskip}
Rev. & $N_{\rm H}$$^{a}$ & $\Gamma$ & $E_{\rm cut}$ & $E_{\rm fold}$ & $E_{\rm cycl}$ & $\sigma_{\rm cycl}$ & $\tau_{\rm cycl}$ & $C_{\rm ISGRI}$ &  $C_{\rm JEMX1}$  & $C_{\rm JEMX2}$  &  $F_{\rm 1-10~keV}$ & $F_{\rm 4-20~keV}^c$  & $F_{\rm 20-50~keV}$  &  $\chi^2_{\rm red}$/d.o.f. \\
   &   &  & keV & keV & keV & keV &  &  &  &  & \multicolumn{3}{c}{(10$^{-10}$\,erg\,cm$^{-2}$\,s$^{-1}$)} & \\
\noalign{\smallskip}
1145-6 &  14.0$^{+1.9}_{-1.7}$ & 0.5$\pm$0.2 & 11.0$\pm$0.8 & 6.2$^{+0.9}_{-0.6}$ & 21.9$^{+0.9}_{-1.2}$ & 3.0$^b$ & 8.1$^{+2.5}_{-2.1}$ & 1.2$^{+0.5}_{-0.4}$ & 1.0$\pm$0.1 & 1.3$\pm$0.1 & 2.6 & 6.5 & 1.2 & 1.0/173 \\
\noalign{\smallskip}
1145-6$^{N}$ & 14.6$^{+1.8}_{-1.7}$ & 0.5$\pm$0.2 & 10.8$\pm$0.6 & 5.7$^{+0.6}_{-0.5}$ & --- & --- & ---  & 1.0$\pm$0.2 & 1.0$\pm$0.1 & 1.3$\pm$0.1 & 2.6 & 6.6 & 1.2 & 1.5/175 \\
\noalign{\smallskip}
1147$^{N}$ &  11.1$^{+6.1}_{-5.3}$ & 0.1$\pm$0.6 & 10.6$^{+1.7}_{-1.6}$ & 4.8$^{+1.4}_{-0.9}$ & --- & --- & --- & 0.8$^{+0.5}_{-0.3}$ & 1.4$^{+0.4}_{-0.3}$ & 1.4$^{+0.4}_{-0.3}$ & 1.6 & 5.5 & 0.6 & 1.1/40 \\
\noalign{\smallskip}
\hline
\multicolumn{15}{l}{$^{a}$: The $N_{\rm H}$ is in units of 10$^{22}$~cm$^{-2}$.}\\ 
\multicolumn{15}{l}{$^b$: This parameter was fixed (see text for details).} \\
\multicolumn{15}{l}{$^{N}$: In these cases, no cyclotron feature was included in the fit. $^c$ Absorbed flux.} \\
\end{tabular}
\label{tab:results}
\end{table*} 

\begin{figure}
\centering
  \includegraphics[width=\columnwidth]{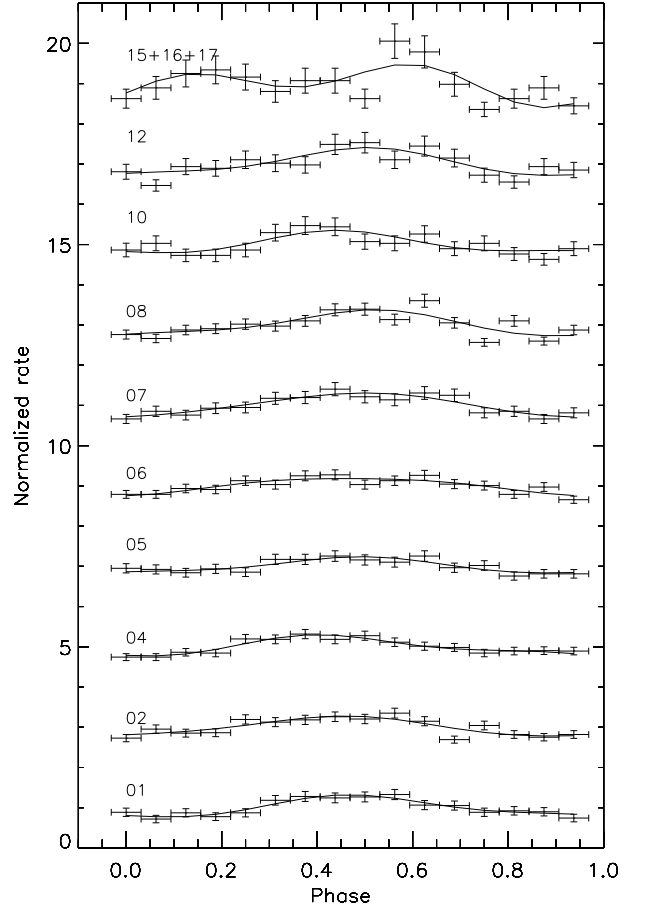}                
  \caption{Normalized pulse profiles of the XRT observations (not phase-connected). The XRT observation number (ID) are indicated by the two last digits for each pulse profile. 
  Each pulse has been divided by its average and vertically 
  displaced by a constant value for clarity. Alignment in the figure is such that in each case the phase of the first Fourier component 
  is 0.5. The solid lines are obtained by the Fourier decomposition using the first two components.}
\label{fig:pulses}
\end{figure}

We performed a detailed timing analysis of the XRT data. 
Pulsations were detected at a period of $\simeq$11.82~s in all observations of a 
sufficiently long duration. In each of these observations, we measured the spin period by using $Z^2$ statistics \citep{buccheri1983} and determined 
the associated uncertainty  using a Monte-Carlo approach, where we simulated 1000 event files with the average modulation of the data and repeated 
the $Z^2$ search on each of them. As detailed in the appendix, we estimate the reliability of the observation by setting a threshold confidence 
level (c.l.) at 99\%.  Only four observations were characterized by a sufficient high quality to permit a reliable determination of the spin period and its 
associated uncertainty. In the others, the spin period was determined at only a low confidence level and we were unable to reliably estimate the corresponding 
uncertainty (the folded lightcurve is reported for each observation in Fig.~\ref{fig:pulses} to allow an easier comparison).  
For each XRT observation, we also computed the fractional root mean square (rms) and its uncertainty using the same Monte Carlo approach described above 
(we simulated 1000 pulse profiles for each observation, see Table~\ref{tab:swifttiming}). 
No significant variation in the rms with luminosity was found. All measurements are compatible with an rms value of 20\%. 

\begin{figure}
\centering
  \includegraphics[width=\columnwidth]{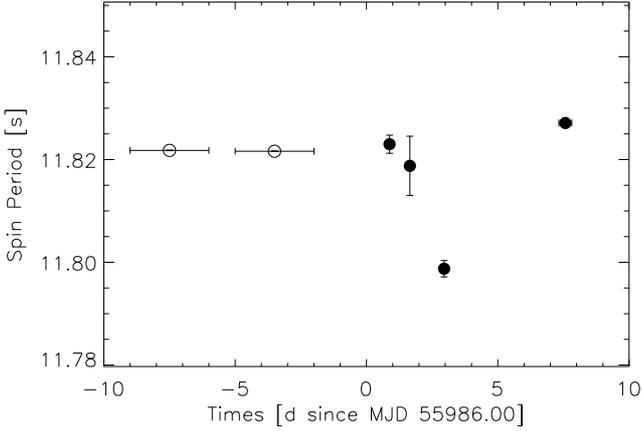}                
  \caption{Spin periods determined using the $Z^2$ statistics. Filled symbols are the subset of the determinations reliable at 99\% c.l. obtained in the present work, 
  open symbols are the determinations of \citet{finger2012} obtained using Fermi/GBM.}
\label{fig:periods}
\end{figure}

All the results of the timing analysis are summarized in Table~\ref{tab:swifttiming}, where we report the uncertainties associated with each 
spin-period measurement (1$\sigma$ c.l.) and a flag expressing the reliability of the measurement. 
Reliable periods from XRT, together with previous determinations from Fermi/GBM (see Sect.~\ref{sec:intro}), are plotted in Fig.~\ref{fig:periods}. 
The first two determinations (Observations 00032293001 and 00032293002) gave a value of the spin period compatible with that determined from the GBM data, 
whereas the spin-period values measured from observations 00032293004 and 00032293008 differed significantly from that.  
As discussed in Appendix~\ref{sec:appendix}, the origin of this deviation cannot be firmly established at present, thus these two 
measurements should probably be taken with caution. 

The XRT monitoring showed that the source began a relatively rapid decay in its X-ray flux after the initial brightening, 
and became fainter by a factor of $\sim$40 in about 22~days. On MJD~56007.78, the source was too faint to obtain meaningful spectral measurements  
with the available exposure times of the XRT pointings (1-2~ks), thus no further observations were performed.   
\begin{table}
\centering
\scriptsize
\caption{Spin period measurements from the XRT data.
Uncertainties are at the 1$\sigma$ confidence level} 
\begin{tabular}{@{}llllll@{}}
\hline
\hline
\noalign{\smallskip}
OBS$^a$ & START TIME & DURATION & $P_{\rm spin}$   & Flag$^b$ &Fractional\\
&(MJD)      &   (days) &  (sec)   &  & rms\\
\noalign{\smallskip}
\hline
\noalign{\smallskip}
01  & 55986.877 & 0.037 & 11.8230$\pm$0.0018 & Y & 	 0.20$\pm$0.03 \\
02  & 55987.653 & 0.012 & 11.819$\pm$0.006 & Y & 	 0.21$\pm$0.02 \\
04$^c$  & 55988.957 & 0.040 & 11.7988$\pm$0.0016 & Y & 	 0.19$\pm$0.02 \\
05  & 55989.395 & 0.074 & 11.7860 & N & 	 0.17$\pm$0.03 \\
06  & 55990.632 & 0.110 & 11.8053 & N & 	 0.18$\pm$0.03 \\
07  & 55992.261 & 0.139 & 11.8151 & N & 	 0.24$\pm$0.03 \\
08$^c$  & 55993.572 & 0.247 & 11.8271$\pm$0.0003 & Y & 	 0.29$\pm$0.03 \\
10  & 55997.219 & 0.008 & 11.81 & N & 	 0.25$\pm$0.04 \\
12  & 55999.367 & 0.335 & 11.8447 & N & 	 0.31$\pm$0.05 \\
15+16+17  & 56006.550 & 1.243 & 11.81 & N & 	 0.46$\pm$0.07 \\
\noalign{\smallskip}
\hline
\multicolumn{6}{l}{$^a$: Only the last two digits of the observation ID.000322930** are shown.} \\
\multicolumn{6}{l}{$^b$: The flag ``Y'' indicates a spin-period determination that is reliable}\\
\multicolumn{6}{l}{at more than 99\% c.l., ``N'' otherwise. }\\
\multicolumn{6}{l}{$^c$: To be considered with caution, see Appendix~\ref{sec:appendix}.} 
\end{tabular}
\label{tab:swifttiming}
\end{table}

\begin{figure}
\centering
$
\begin{array}{c|c}
  \includegraphics[width=4.2cm]{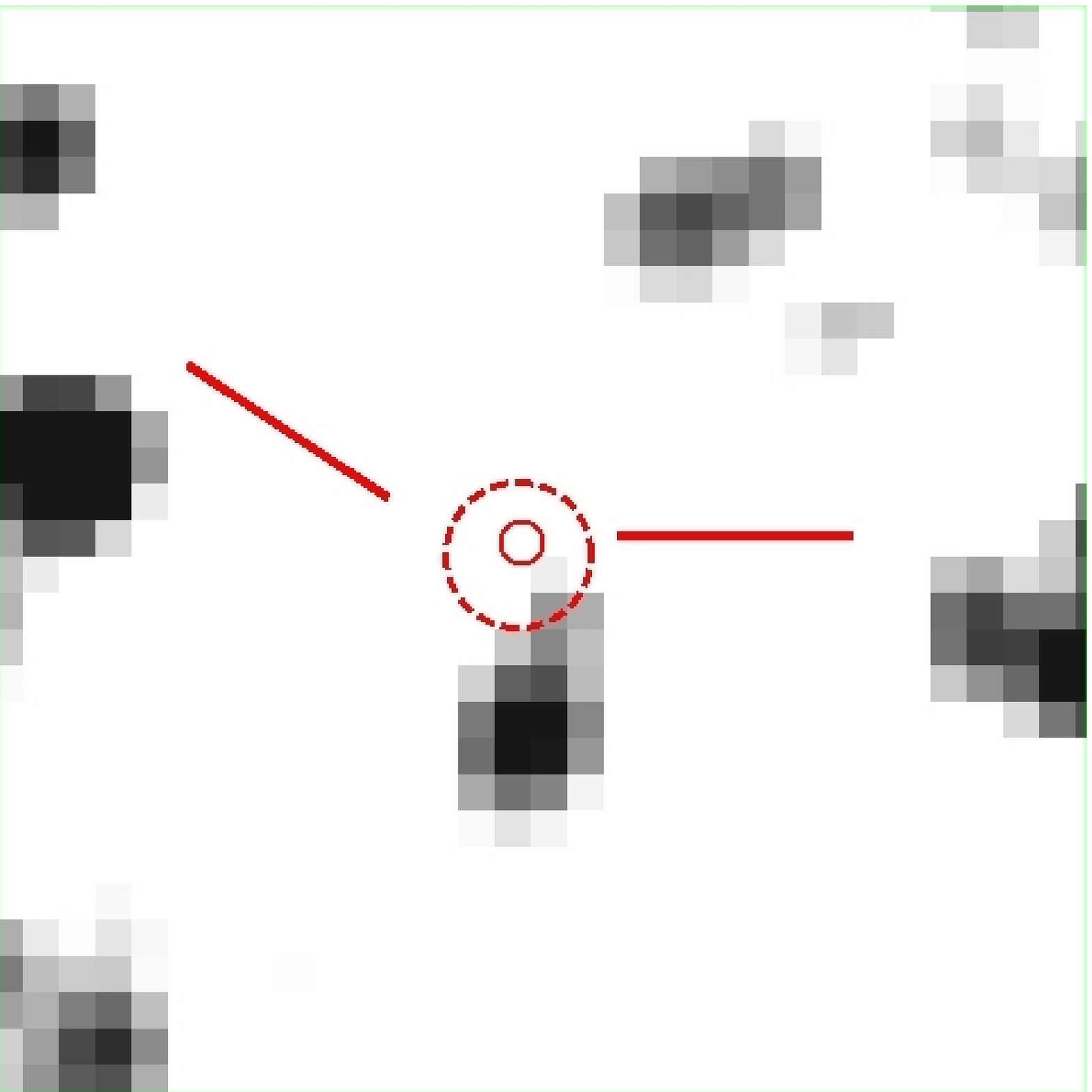} &  \includegraphics[width=4.2cm]{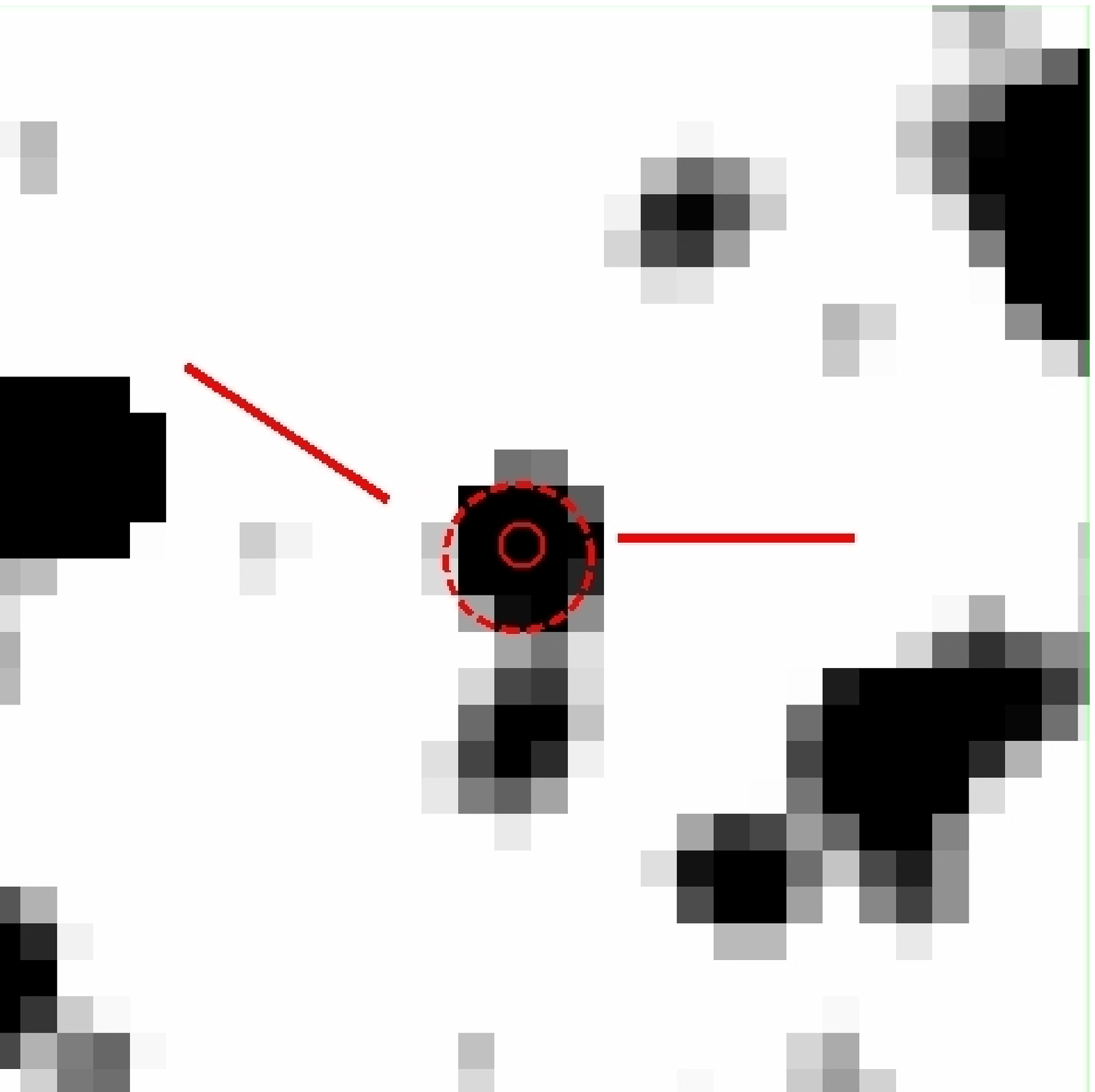} \\
\end{array}
$  
  \caption{Images of the field of view around \igr\ from the 2MASS archive (4\arcmin$\times$4\arcmin\ ) \emph{Left:} image in the J-band together with 
  the \swift\ (dashed circle) and \chandra\ (solid circle) localizations of \igr.\ The solid lines represent the localization of the  
  proposed IR counterpart to \igr,\  2MASS $J18175218-1621316$, which lies within the X-ray error circles. 
  \emph{Right:} Same as previous figure but for the K-band, in which the counterpart is detected.}
\label{fig:2masscounterpart}
\end{figure}

\section{Counterparts}
\label{sec:counterparts}

The arcsecond localization of \igr\ obtained first through the \swift\ observations and then refined with \chandra\ (see Sect.~\ref{sec:intro}), 
permitted to search for possible associated counterparts to the source in the optical and IR domains. 
A search in the 2MASS catalog \citep{2mass} revealed that 
the object 2MASS\,$18175218-1621316$ is the most likely IR counterpart of the X-ray pulsar 
\citep[see also][]{li12,paizis12}. We verified that the 2MASS object has been detected in K band with a magnitude of $K=13.14\pm0.04$, 
while in the J and H bands only upper limits were reported (magnitudes of 16.7 and 15.9, respectively). 
No obvious counterpart could be found at the position of \igr\ from the USNO B-1 catalog \citep{usno}. 
At present, the lack of any detailed information in the J and H bands, combined with the unknown distance to the source, does not 
allow us to significantly constrain the nature of the companion to \igr.\

\section{Discussion and conclusions}
\label{sec:discussion}

\igr\ was detected for the first time with \inte\ during the period ranging from 
55\,986.1~MJD to 55\,992.5~MJD, even though Fermi/GBM data indicated that the outburst of the source 
might have already started on 55977~MJD. A follow-up monitoring with XRT was initiated shortly after the discovery and continued 
until 56\,007.8~MJD, measuring a decay in the source X-ray flux of a factor of $\sim$40 (from 2.9$\times$10$^{-10}$\,\ferg to 8$\times$10$^{-12}$\,\ferg) 
in $\sim$22~days. 
The detection of pulsations at $\simeq$11.82~s with the GBM and XRT, 
led soon to classifying \igr\ as an accreting X-ray pulsar \citep{halpern12}. 

The spectral analysis of all the available \inte\ and \swift\ data collected during the outburst 
revealed that the X-ray emission from the source was heavily obscured. We estimated an absorption column density  of
about an order of magnitude higher than the Galactic value expected in that direction 
\citep[$N_{\rm H}=1.6\times$10$^{22}$~cm$^{-2}$][]{dickey90}. In the brightest phase of the outburst (from MJD~55\,986.1 to MJD~55\,990.8), 
the broad-band XRT/JEM-X/ISGRI spectrum could be well fit with a cut-off power-law model ($\Gamma=0.5\pm0.2$, 
$E_{\rm fold}=(5.7\pm0.6)$~keV, $E_{\rm cut}=(10.8\pm0.6)$~keV), 
but some significant residuals were left above 10~keV. Following \citet{turler12}, 
we showed in Sect~\ref{sec:swift} that the fit could be improved by adding a cyclotron absorption feature 
with a centroid energy at ($22\pm1$)~keV. The feature was not significantly detected during the latest stages of the outburst 
(from MJD~55\,991 onward), most likely owing to the relatively low S/N of the data. 
As the centroid energy of the line lies in the energy range in which the JEM-X and ISGRI instrument  responses 
overlap, we caution that at present it is impossible to firmly exclude that some calibration uncertainties might have affected 
this detection. Future observations are thus needed to establish its presence. This feature 
would however not be unexpected in the case of an X-ray pulsar. 

A hard spectrum with $\Gamma<0.6$ is sometimes observed from sources that display cyclotron lines \citep{coburn2002}, and could be    
related to the effects of cyclotron scattering \citep{ferrigno09}.
Cyclotron resonant-scattering features (hereafter CRSFs) might also appear in the spectra of these objects in the form of absorption 
lines. The latter are caused by resonant scattering of photons off the electrons in Landau levels 
in the strong magnetic field of the neutron star (10$^{11}$-10$^{13}$~G). 
The centroid energy of the fundamental absorption feature is related 
to the strength of the magnetic field in the scattering region by the equation \citep[see e.g.,][]{wasserman1983} 
$E_{\rm cycl}$$\simeq$11.6/(1+z)(B/10$^{12}$~G)~keV, 
where $B$ is the magnetic field strength in  gauss and $z$+1=1.31 is the gravitational redshift (we used the canonical 
neutron-star mass and radius of 1.4~M$_{\sun}$ and 10~km, respectively). 
If the cyclotron feature is confirmed in the future, we will be able to infer for \igr\ a magnetic field strength 
of $\sim$2$\times$10$^{12}$~G. 

A similar magnetic field strength, associated with pulsations at $\sim$12~s, would make \igr\ a likely Be X-ray binary candidate \citep[hereafter 
BeXBs; see e.g.][for a review]{reig11}. According to this interpretation, the Corbet diagram \citep{corbet86} suggests for this source an orbital 
period in the range 20-50~days. Despite this apparently straightforward association, such a relatively short orbital 
period makes the interpretation of the event recorded by \inte\ and \swift\ in terms of the so-called BeXB ``type-I'' outbursts challenging  
\citep[see e.g.,][]{reig11}. As the latter occur almost regularly at each periastron passage in these systems, previous detections of \igr\ 
should probably be expected during the long-term monitoring of the sky around the source performed so far with \inte.\ 
By using the online tool {\sc HEAVENS}\footnote{http://www.isdc.unige.ch/heavens/}, we checked that the total effective exposure-time at the coordinates of \igr\ is 2.4~Ms for ISGRI and 410~ks for JEM-X (considering all publicly available data from 2003 February 28 to 2010 March 1). 
A search in these archival data did not result in any previous significant detection of the source, and we derived a 5$\sigma$ upper limit to the 
source flux of 0.5~mCrab in the 17-40~keV energy band and 1.7~mCrab in the 3-10~keV energy band. 
The interpretation of the event recorded from \igr\ as a rare BeXB type-II outburst \citep{stella86}, might also be questionable.
The X-ray luminosity reached during these events is typically on the order of $L_{\rm X}$$\simeq$10$^{38}$~erg~$s^{-1}$, and would thus imply an unrealistically large 
 distance for \igr\ ($\gg$10~kpc, using the peak flux measured in Sect.~\ref{sec:swift}). 
If the event reported here is the first episode of intense X-ray activity ever displayed by \igr,\ this source could be a further example of those 
``dormant'' X-ray binaries that spent most of their time life in a quiescent state before suddenly changing into bursting X-ray sources 
\citep[see e.g. the case of RX\,J0440.9$+$4431;][]{tsygankov2012,usui2012}. 
 
At odds with typical BeXBs, \igr\ appears to be characterized by a peculiar high absorption column-density. 
A value of $N_{\rm H}$$\simeq$10$^{23}$~cm$^{-2}$ is more commonly observed in the so-called highly obscured high-mass X-ray binaries 
\citep[HMXBs; see e.g.][for a review]{chaty10}. A case of a highly absorbed HMXB displaying a cyclotron scattering feature is the 
persistent source GX~301$-$2 \citep[see][and references therein]{suchy2012}.
In these sources, the neutron star is accreting matter from the intense wind of its massive 
companion giving rise to a fairly persistent X-ray luminosity in the range 10$^{34}$-10$^{36}$~erg~s$^{-1}$ 
(the exact value depends mainly on the strength of the wind and on the orbital separation between the supergiant and the neutron star). 
Bright short flares, lasting a few hours and reaching a peak flux $\sim$10-50 higher than the persistent emission, are sporadically observed 
from these objects and ascribed to episodes of accretion from high density ``clumps'' of stellar wind material 
\citep{zand05, bozzo08, krey08, negueruela10, bozzo11}. 
The dynamic range of the X-ray flux measured from \igr\ (taking into account the upper limit derived above), might thus still be 
marginally compatible with that measured for the highly obscured HMXBs, although the scale time of the outburst decay inferred from the \swift\ monitoring 
($\sim$22~days) seems hardly reconcilable with that expected for an episode of clumpy wind accretion. 

Even though some of the results from the timing analysis of the XRT data have to be interpreted with caution (see \ref{sec:appendix}), 
in Sect.~\ref{sec:data} we reported the possible detection of a spin-period variation from the source. If this is interpreted in terms 
of an accretion torque, the average spin-up and spin-down rates inferred from a linear fit to the measurements before and after MJD~55\,989.13 
(see Fig.~\ref{fig:periods}) would be on the order of $\sim$$10^{-7}$\,s/s, 
i.e. compatible with those measured in a number of accreting high-mass X-ray binaries \citep[see also the case of GX\,1-4]{bildsten1997,ferrigno2007}. 
We note that similar changes in the spin period of the source might also be due to the orbital modulation, but the limited number of 
currently available estimates of $P_{\rm spin}$ do not allow us to investigate further this possibility. 
Future observations of the source in outburst are needed to strengthen and confirm any detection of  
spin-period variations in \igr.\ 

As an alternative interpretation, we also suggest that \igr\ might be another example of those peculiar binaries displaying a complex X-ray variability 
with intermediate properties 
in-between those of young wind-accreting systems and more evolved disk accretors. The latter can display prolonged periods of quiescence and undergo moderate to bright 
outbursts owing to instabilities in the accretion disk \citep[see e.g.,][and references therein]{lasota}. Among the known peculiar disk-accreting systems, a  
particularly relevant case here is Her~X-1, which is also known to display a cyclotron absorption line at $\sim$40~keV 
\citep[see e.g.,][and references therein]{vasco11}. 
Spectroscopic observations of \igr\ in other energy domains (e.g. IR) might help in clarifying the nature of its companion star and unveil the real nature 
of the mechanism regulating the X-ray variability displayed by this new \inte\ transient.  

\begin{acknowledgements} 
We thank the \swift\ team for the 
prompt scheduling of all the follow-up 
observations performed after the discovery of 
\igr.\ We acknowledge P. Kretschmar and an anonymous referee for carefully 
reading the manuscript and giving valuable comments.
\end{acknowledgements}

\bibliographystyle{aa}
\bibliography{igrj18179_f}

\appendix

\section{Spin-period determination}
\label{sec:appendix} 

\begin{figure}
\centering
  \includegraphics[width=\columnwidth]{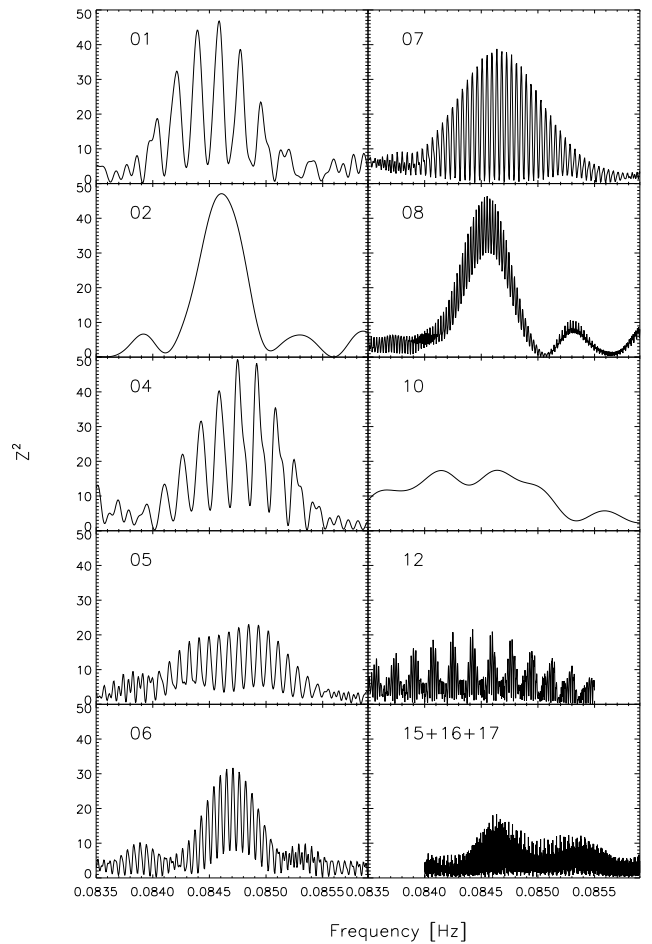}                
  \caption{Periodograms obtained from the $Z^2$ statistics applied to the XRT observations reported in Sect.~\ref{sec:swift}. 
  The OBSIDs are indicated by the two last digits in each panel.}
\label{fig:z2}
\end{figure}

The uncertainty and the significance of the pulse period derived using the $Z^2$ 
statistics in each of the XRT observations reported in Sect.~\ref{sec:swift}, could be reliably estimated only by performing Monte Carlo 
simulations, as the individual determinations obtained by  
scanning frequencies from a single event file are not independent. 
Furthermore, we verified that the formal significance of the $Z^2$ peak is not a good estimate of the reliability 
of the measurement for the \swift\ observations. The latter comprise several relatively short ($\la$1\,ks) snapshots 
containing only a few hundreds photons each and, as a result, the corresponding periodogram has the shape of a wide-peak   
(the diffraction figure of the window function) with evident interference fringes produced by the fragmentation of the observations 
(see Fig.~\ref{fig:z2}). 
\begin{figure}
\centering
  \includegraphics[width=\columnwidth]{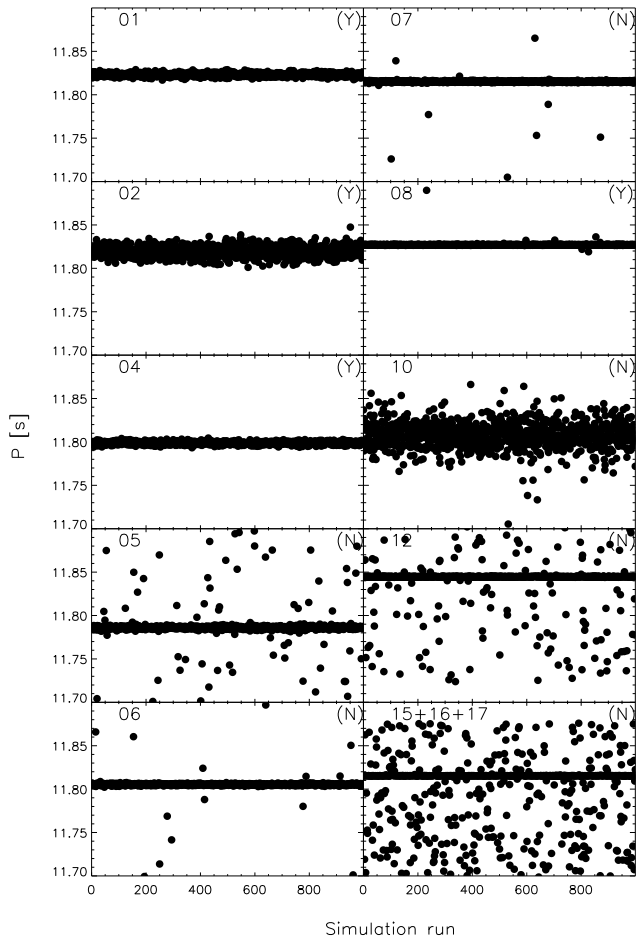}                
  \caption{Spin periods derived from the simulated event files for each of the XRT observation reported 
  in Sect.~\ref{sec:swift}. The OBSIDs are indicated with the two last digits for each panel.}
\label{fig:simulations}
\end{figure}

To perform the Monte Carlo simulations, we determined in each observation the best pulse period and approximated the corresponding pulse profile by   
using its first two Fourier components (a check was performed {\it a posteriori} to verify that this description is accurate in each case by inspecting 
the single pulse profiles; see Fig.~\ref{fig:pulses}). Event files with an average event rate equal to that measured from the source were then simulated in the 
good time intervals (GTIs) of the observations by introducing two sinusoidal modulations at the estimated spin period and its first harmonic. 
The amplitudes of the modulations were assumed to be the same as those measured from the Fourier decomposition of the real source pulse profiles. 
In each of the simulated event files, the pulse period was then measured with the $Z^2$-statistics method and recorded. 
We report the results of this analysis in Fig.~\ref{fig:simulations}. 

The spin periods determined from the simulated files cluster around the real value measured in the corresponding observations and follow roughly a Gaussian distribution. 
The shape of the Gaussian could in principle be used to estimate in each case the uncertainty in the spin-period determination, but this procedure is complicated in all 
cases by a certain number of ''outliers''. 
The number of these points, whose ''wrong'' value of the spin period is due to the sporadic displacement of the peak in the 
$Z^2$-statistics periodogram at the moment of the best period determination, is larger for shorter observations and observations characterized by a 
smaller number of source events. In \citet{bozzo11}, we argued that the number of outliers with respect to the total number of measurements could give an estimate 
of the reliability (significance) of the spin-period detection in an observation.  

To identify the outliers in the present case, we: (i) collected all the spin periods determined from the 90\% of the simulations that 
gave periods closer to those measured from the real data (''central spin periods''), (ii) estimated the average spin period 
and the variance $\sigma^2$ of the corresponding Gaussian distribution, and (iii) identified all the remaining realizations that gave 
a spin period differing by more that 2.6$\sigma$ from the averaged one. 
We rejected as ''not-reliable'' the XRT observations in which the number of outliers were found to significantly exceed \citep[99\% c.l., taking into account also the 
intrinsic uncertainty in $\sigma$ and the Poissonian nature of the key variables; see e.g.,][]{sheskin} the value expected 
from the Gaussian distribution of the simulated central spin periods.   

As a final remark, we note that, in all cases, the timing analysis was performed on XRT event files that were uncorrected for pile-up (see sect.~\ref{sec:data}), 
as the latter is known to affect mostly the spectral energy distribution of photons recorded from the 
source\footnote{See http://www.swift.ac.uk/analysis/xrt/pileup.php}. We checked {\it a posteriori} that an analysis similar to that described above performed on the 
pile-up corrected XRT event files of the reliable spin-period determinations in observations 00032293001 and 00032293002 would give values fully 
in agreement (to within the uncertainties) with those reported in Table~\ref{tab:swifttiming}. For the observations 00032293004 and 00032293008, the 
lower flux and/or number of counts led, in contrast to an unreliable spin-period determination once the corresponding data are corrected for pile-up. 
As the effect of pile-up on these observations cannot be checked further owing to the limited statistics, the corresponding values of the spin periods reported 
in Table~\ref{tab:swifttiming} should be interpreted with caution.

\end{document}